\documentclass{iau}
\usepackage{graphicx}

\newcommand{\apj}{{\it ApJ}}
\newcommand{\aj}{{\it AJ}}
\newcommand{\mnras}{{\it MNRAS}}
\newcommand{\aanda}{{\it A\&A}}
\newcommand{\pasp}{{\it PASP}}

\title[IAUS289.~~Classical Cepheids, what else?] 
{Classical Cepheids, what else?}

\author[G. Bono et al.]   
{G. Bono,$^{1,2}$ L. Inno,$^{1,3}$ N. Matsunaga,$^4$ K. Genovali,$^1$ B. Lemasle,$^5$\\
F. Primas,$^3$ \and M. Romaniello$^3$}

\affiliation{$^1$Dipartimento di Fisica, Universit\`a di Roma Tor Vergata,
Via della Ricerca Scientifica 1, 00133 Roma, Italy\\ email: {\tt bono@roma2.infn.it}\\
$^2$INAF--OAR, via Frascati 33, 00040 Monte Porzio Catone, Italy\\[\affilskip]
$^3$ESO, Karl-Schwarzschild-Str. 2, 85748 Garching bei M\"unchen, Germany\\[\affilskip]
$^4$Department of Astronomy, School of Science, The University of Tokyo, 7-3-1 Hongo, Bunkyo-ku, Tokyo 113-0033, Japan\\[\affilskip]
$^5$Sterrenkundig Instituut `Anton Pannekoek,' University of Amsterdam,\\ Science Park 904, P. O. Box 94249, 1090 GE Amsterdam, The Netherlands}

\pubyear{2013}
\volume{289}  
\jname{Advancing the Physics of Cosmic Distances}
\editors{Richard de Grijs, ed.}

\begin{document}

\maketitle

\begin{abstract}
We present new and independent estimates of the distances to the
Magellanic Clouds (MCs) using near-infrared (NIR) and optical--NIR
period--Wesenheit (PW) relations.  The slopes of the PW relations are,
within the dispersion, linear over the entire period range and
independent of metal content.  The absolute zero points were fixed
using Galactic Cepheids with distances based on the infrared
surface-brightness method.  The true distance modulus we found for the
Large Magellanic Cloud---$(m-M)_0 = 18.48 \pm 0.01 \pm 0.10$ mag---and
the Small Magellanic Cloud---$(m-M)_0 = 18.94 \pm 0.01 \pm 0.10$
mag---agree quite well with similar distance determinations based on
robust distance indicators.  We also briefly discuss the evolutionary
and pulsation properties of MC Cepheids.  

\keywords{distance scale, Magellanic Clouds, stars: evolution, stars: pulsations}
\end{abstract}

\firstsection 
\section{Introduction}

The modern use of classical Cepheids as primary distance indicators
dates back more than half century, and in particular to \cite[Baade
  (1948)]{baade48} and \cite[Hubble (1953)]{hubble53}.  However, for
the first detailed discussions regarding the universality of the
period--luminosity (PL) relation of classical Cepheids, we refer to
the seminal papers by \cite[Sandage (1962)]{sandage62},
\cite[Gascoigne \& Kron (1965)]{gasc65}---Magellanic Cloud (MC)
Cepheids---\cite[Kayser (1967)]{kayser67}---NGC~6822 Cepheids---and
\cite[Sandage \& Tammann (1968)]{sandage68}.  From a theoretical point
of view, the first detailed evolutionary investigations date back to
\cite[Kippenhahn \& Smith (1969)]{kippenhahn69}, \cite[Iben \& Tuggle
  (1975)]{iben75}, and \cite[Becker et al. (1977)]{becker77}. Linear,
non-adiabatic pulsation models for classical Cepheids were developed
by \cite[Cox (1979)]{cox79}, \cite[Iben (1974)]{iben74}, and
\cite[Castor (1971)]{castor71}.  The pioneering modeling of nonlinear,
radiative Cepheid models dates back to \cite[Christy (1968,
  1975)]{christy68,christy75} and \cite[Stobie (1969)]{stobie69}.  The
main outcome of both theoretical and empirical investigations is that
classical Cepheids are robust primary distance indicators, since they
obey a universal optical PL relation. This evidence is also supported
by optical PL--color (PLC) relations for Galactic (\cite[Fernie
  1967]{fernie67}; \cite[Sandage \& Tammann 1969]{sandage69}), Large
and Small Magellanic Cloud (LMC, SMC: \cite[Butler 1978]{butler78};
\cite[Feast \& Balona 1980]{feast80}; \cite[Caldwell \& Coulson
  1987]{caldwell87}) Cepheids.  However, empirical and theoretical
investigations already suggested that the use of PLC relations is
hampered by uncertainties affecting the reddening corrections
(\cite[Stift 1982]{stift82}).

Observational scenarios regarding the use of classical Cepheids as
distance indicators were further enriched by the seminal near-infrared
(NIR) investigations of Magellanic and Galactic Cepheids by
\cite[Laney \& Stobie (1986, 1993, 1994)]{laney86,laney93,laney94}.
The main advantages of using mid-IR PL relations is that they are
minimally affected by uncertainties in reddening, nor by the intrinsic
width in temperature of the instability strip (\cite[Bono \&
  Stellingwerf 1993]{bono93}).  The use of NIR photometry also
provided a substantial improvement in the precision of individual
distances based on the infrared surface-brightness (IRSB) method
(\cite[Gieren 1989]{gieren89}; \cite[Barnes \& Evans 1976]{barnes76};
\cite[Welch 1994]{welch94}; \cite[Groenewegen 2004]{groe04};
\cite[Storm et al. 2011a,b]{storm,stormb}; and references therein).
However, the quantum jump in the use of classical Cepheids as standard
candles arrived with the advent of microlensing experiments. During
the last dozen years, the number of known regular variables, and in
particular RR Lyrae and Cepheids, both in the Galaxy and the MCs,
increased by more than an order of magnitude ({\sc macho}:
\cite[Alcock et al. 2000]{alcock00}; {\sc eros}: \cite[Marquette
  1999]{marquette99}; {\sc ogle}: \cite[Soszy{\'n}ski et
  al. 2012]{sos2012}).

The modeling of the nonlinear behavior of Cepheids was placed on a
solid basis thanks to the coupling between hydrodynamical equations
and time-dependent convection (\cite[Stellingwerf 1982,
  1984]{stel82,stel84}; \cite[Bono \& Stellingwerf 1993]{bono93};
\cite[Buchler et al. 1990]{buchler90}; \cite[Keller \& Wood
  2006]{keller06}; \cite[Marconi et al. 2010]{marconi10};
\cite[Fiorentino et al. 2012]{fiorentino2012}). Comparisons between
convective models and new results from photometric surveys indicated
that optical and NIR PL relations might not be universal.  This opened
up a lively debate from both theoretical (\cite[Alibert et
  al. 1999]{Alibert99}; \cite[Bono et al. 2000, 2010]{bono00,bono10};
\cite[Baraffe \& Alibert 2001]{baraffe01}) and empirical
(\cite[Sandage et al. 2004]{sandage04}; \cite[Sakai et
  al. 2004]{sakai04}; \cite[Ngeow et al. 2005, 2008]{ngeow05,ngeow08};
\cite[Fouqu\'e et al. 2007]{fouque2007}; \cite[Groenewegen
  2008]{groe08}; \cite[Storm et al. 2011a,b]{storm,stormb}; \cite[Inno
  et al. 2013]{inno13}) points of view.

The two crucial issues addressed in these investigations are (i) the
dependence on metallicity of both the slope and the zero point of the
PL and PLC relations and (ii) the linearity of the PL and PLC
relations over the entire period range. In particular, recent
spectroscopic and photometric studies indicate that metal-poor
Cepheids are, at fixed period, brighter than their metal-rich
counterparts, and that the PL relation is not linear (\cite[Marconi et
  al. 2005]{marconi05}; \cite[Ngeow et al. 2005]{ngeow05};
\cite[Romaniello et al. 2008]{romaniello08}).  However, no general
consensus has yet been reached as regards the metallicity dependence,
and indeed suggestions have been made of either a marginal dependence
(\cite[Gieren et al. 2005]{gieren2005}) or an inverse trend,
i.e. metal-poor Cepheids are, at fixed period, {\em fainter} than
metal-rich ones (\cite[Macri et al. 2006]{macri06}).

However, the use of period--Wesenheit (PW) relations appears very
promising. The Wesenheit indices are pseudo-magnitudes, and their key
property is that they are reddening-free (\cite[Madore
  1982]{madore82}).  Recent theoretical and empirical investigations
indicate that both NIR and optical--NIR PW relations are independent
of metal content and linear over the entire period range
(\cite[Groenewegen 2008]{groe08}; \cite[Matsunaga et
  al. 2011]{matsunaga11}; \cite[Storm et al. 2011a,b]{storm,stormb}).

\section{NIR data of MC Cepheids} 

The single-epoch {$J,H,K_{\rm}$} data for 3042 LMC (1840 fundamental
[FU], 1202 first-overtone [FO]-mode variables) and 4150 SMC (2571 FU,
1579 FO) Cepheids were taken from the NIR catalog of the {\sl
  IRSF}/SIRIUS NIR MC Survey of \cite[Kato et al. (2007)]{kato07} and
transformed to the {\sc 2mass} NIR photometric system.  The $V$ and
$I$ mean magnitudes, the $V$-band amplitude, the period, and the
pulsation phase for the same Cepheids were extracted from the {\sc
  ogle iii} catalog by \cite[Matsunaga et al. (2011)]{matsunaga11}.
We transformed the single-epoch NIR magnitudes by adopting a template
light curve (\cite[Soszy{\'n}ski et al. 2005)]{sos05}.  For 41
long-period Cepheids in the LMC, we adopted the mean magnitudes from
\cite[Persson et al. (2004)]{persson04}. The entire data set is
discussed in detail by \cite[Inno et al. (2013)]{inno13}.  Fig.~1
shows the period distribution for LMC and SMC FU and FO Cepheids. The
main differences between the distributions in the two galaxies are
owing to the evolutionary and pulsation properties of the Cepheids
(see Section~3).

\begin{figure}[htb!]
\vspace*{-3.5 cm}
\begin{center}
\includegraphics[height=0.75\textheight,width=0.75\textwidth]{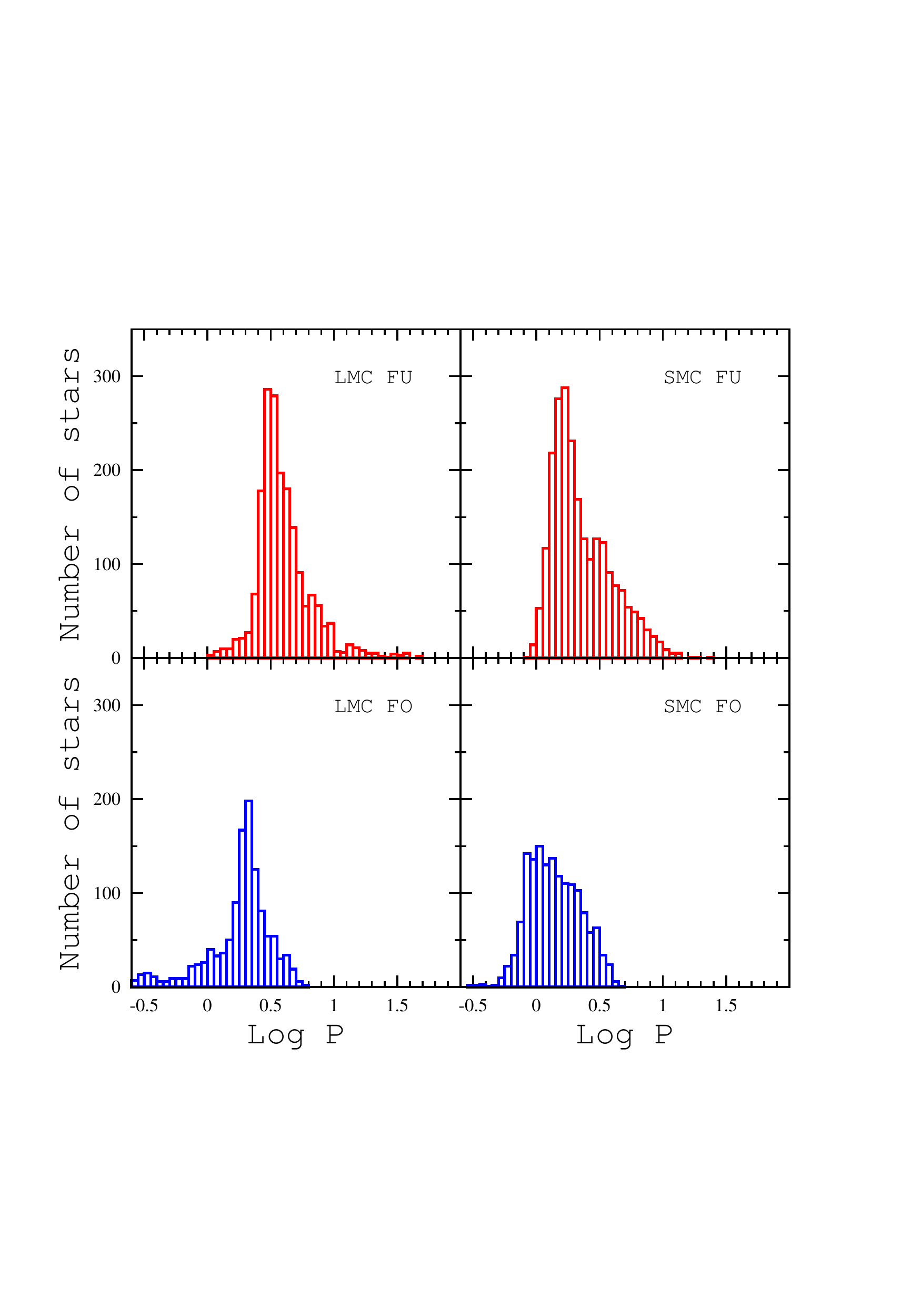} 
\vspace*{-2.9 cm}
\caption{Period distributions for FU (top) and FO (bottom) LMC (left) and SMC (right) 
Cepheids.}\label{fig1}
\end{center}
\end{figure}

\section{Evolutionary and pulsation properties of MC Cepheids} 

The period distributions of FU (top) and FO (bottom) LMC and SMC
Cepheids plotted in Fig.~1 show two well-known differences:
\begin{enumerate}
\item The peak in the period distribution of FU and FO SMC Cepheids is
  found at shorter periods compared with LMC Cepheids. In the former
  system, the peak is located at $\log P \sim$ 0.2 (FU) and $\sim$ 0.0
  [days], while in the LMC, the peak is found at $\log P \sim$ 0.5
  (FU) and $\sim$ 0.3 [days].

\item The period distribution of SMC Cepheids is broader than that of
  LMC Cepheids, and indeed in the former system the long-period tails
  of both the FU and FO Cepheids exhibit shallower profiles.
\end{enumerate}

The intrinsic difference in the period distribution between LMC and
SMC Cepheids, combined with the evidence that SMC Cepheids are, for a
given period, systematically bluer than LMC and Galactic Cepheids, was
noted more than 40 years ago by \cite[Gascoigne (1969, 1974)]{gasc69,
  gasc74}.  On the basis of evolutionary and pulsation models,
\cite[Iben (1967)]{iben67} and \cite[Christy (1971)]{christy71}
suggested that these differences could be explained as a difference in
metal content.  The observational scenario was soundly confirmed by
microlensing experiments (\cite[Sasselov et al. 1997]{sasselov97};
\cite[Soszy{\'n}ski et al. 2012]{sos2012}) and more detailed
theoretical investigations (\cite[Bono et al. 2000]{bono00}).  The
Cepheid period distribution might play a crucial role in constraining
the recent star-formation rate and the properties of young stellar
populations (\cite[Alcock et al. 1999]{alcock99}). However, the
problem is far from trivial, since the period distribution depends not
only on the initial mass function and the most recent star-formation
episodes, but also on the Cepheid metallicity distribution.

Our knowledge of the metallicity distribution of MC Cepheids is quite
limited. Accurate spectroscopic measurements based on high-resolution
spectra are only available for a few dozen relatively bright Cepheids
(\cite[Luck et al. 1998]{luck98}).  On the basis of 22 LMC and 14 SMC
Cepheids, \cite[Romaniello et al. (2008)]{romaniello08} found mean
iron abundances for LMC Cepheids of [Fe/H] = $-0.33$ dex, with
individual abundances ranging from $-0.62$ to 0.10 dex, while for SMC
Cepheids [Fe/H] = $-0.75$ dex on average, with abundances ranging from
$-0.87$ to $-0.63$ dex.


\begin{figure}[htb!]
\begin{center}
\includegraphics[width=0.75\textwidth]{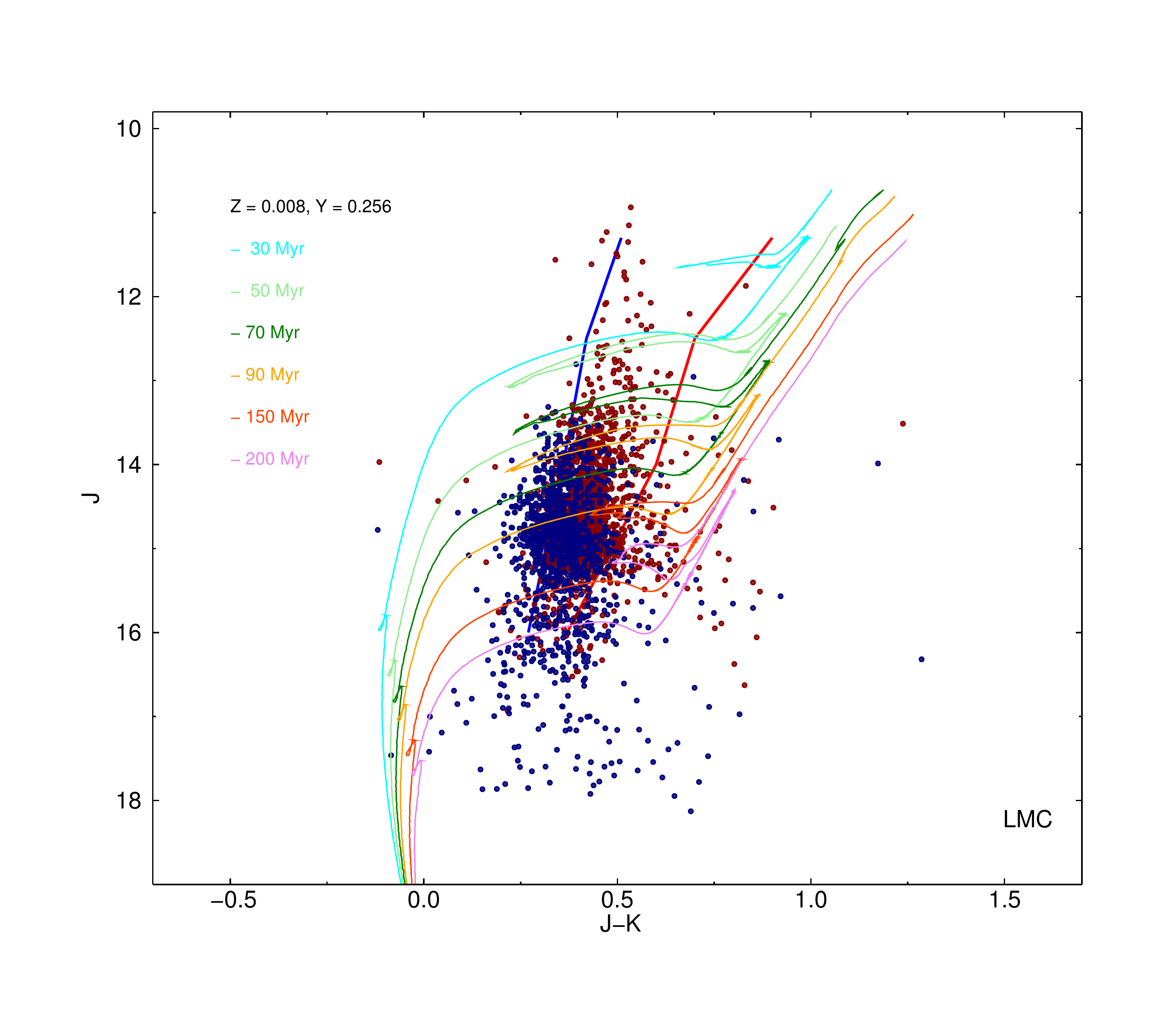} 
\caption{NIR ($J,J-K$) CMD of the FU (red) and FO (blue) LMC
  Cepheids. The differently colored lines display stellar isochrones
  with ages ranging from 30 to 200 Myr (see labels). The isochrones
  are based on non-canonical evolutionary models that include mild
  convective core overshooting during the central hydrogen-burning
  phase and a canonical mass-loss rate ($\eta=0.4$; \cite[Pietrinferni
    et al. 2006]{pietrinferni06}). They were constructed by assuming a
  scaled-solar chemical mixture at fixed metal ($Z=0.008$) and helium
  ($Y= 0.256$) abundances.  The vertical red and blue lines show the
  cool and hot edges, respectively, of the instability strip predicted
  by pulsation models. They account for the modal stability of both FU
  and FO Cepheids and have been constructed by assuming a
  non-canonical mass--luminosity relation and a similar chemical
  composition (\cite[Marconi et al. 2005]{marconi05}).  Theoretical
  predictions were plotted by assuming a true distance modulus of $\mu
  =18.48$ mag a mean reddening of $E(V-I)= 0.09$ mag (\cite[Haschke et
    al. 2011]{haschke11}) based on the reddening law of \cite[Cardelli
    et al. (1989)]{cardelli89}.}\label{fig2}
\end{center}
\end{figure}

To further constrain the evolutionary status of MC Cepheids, Fig.~2
shows the NIR ($J,J-K$) color--magnitude diagram (CMD) of the selected
Cepheids.  Data points plotted in Fig.~2 show that FO Cepheids (blue
circles) attain, as expected, bluer colors compared with FU Cepheids
(red circles).  The colored lines show a set of isochrones with a
scaled-solar chemical mixture and a chemical composition ($Z=0.008,
Y=0.256$) typical of LMC Cepheids (\cite[Pietrinferni et al. 2004,
  2006]{pietrinferni04,pietrinferni06})\footnote{\tt
  http://albione.oa-teramo.inaf.it/}.The isochrones are based on
evolutionary models that account for mild convective core overshooting
during the central hydrogen-burning phase and include a canonical
mass-loss rate ($\eta=0.4$). The comparison between theory and
observations indicates that LMC Cepheids have ages ranging from a few
tens to a couple of hundred Myr. The isochrones in the old ($t\approx
200$ Myr) and the young ($t\approx 30$ Myr) age ranges are
characterized by `blue loops' that do not cover the range in color of
observed Cepheids. This is a well-known problem of intermediate-mass
evolutionary models. The extent in color of the blue loops is affected
by several physical mechanisms and by the input physics adopted.  The
treatment of mixing processes at the edge of the convective core and
at the base of the convective envelope \cite[(Cassisi \& Salaris
  2011)]{cassisi11}, as well as the mass-loss efficiency
(\cite[Matthews et al. 2012]{matthews12}), are most popularly
considered the culprits. However, the interested reader is referred to
\cite[Bono et al. (2000)]{bono00}, \cite[Prada Moroni et
  al. (2012)]{prada12}, and \cite[Neilson et al. (2011)]{neilson11}
for more detailed discussions.  

The vertical blue and red lines show the predicted edges of the
instability strip for a similar chemical composition. The blue and red
lines account for the modal stability of both FU and FO Cepheids
(\cite[Marconi et al. 2005]{marconi05}). Data points included in this
figure reveal that theory and observations agree quite well.  The
predicted blue edge of the instability strip appears to be slightly
cooler than the observed edge. However, the comparison between theory
and observations was performed for fixed chemical composition,
assuming the same reddening and neglecting depth effects. Cepheids
located at fainter magnitudes and outside the instability strip are
typically affected by higher extinction.

The SMC Cepheids show a very similar distribution in the NIR CMD (see
Bono et al. 2000, their fig. 1). The difference in the period
distribution between LMC and SMC Cepheids is explained by the current
evolutionary framework. More metal-poor isochrones are characterized,
at fixed age, by more extended blue loops. This means that the minimum
mass crossing the instability strip decreases when moving from
metal-rich to more metal-poor stellar systems. This difference,
combined with the fact that the lifetime spent inside the instability
strip by low-mass Cepheids is longer than for higher masses, causes an
increase in the relative number of short-period Cepheids.

\section{NIR PW relations} 

On the basis of two magnitudes (e.g., $m_I$ and $m_V$), we can define
a Wesenheit index, e.g., $W(V,I)= m_I - [A_I/E(V-I)]\times (m_V-m_I)$.
We already mentioned that one of the main advantages of using PW
relations for estimating Cepheid distances is that Wesenheit indices
are independent of uncertainties affecting reddening estimates of
Galactic and extragalactic Cepheids.  Once we adopt a reddening law,
the ratio of visual-to-selective absorption---$R_V=A_V/E(B-V)$---and
assuming that the reddening law is universal, we can determine the
color coefficient to define the Wesenheit pseudo-magnitudes.

The main advantage of using NIR measurements of Cepheids, compared
with the use of optical data, is that their pulsation amplitude
decreases with increasing wavelength.  Therefore, estimating Cepheid
mean magnitudes in NIR bands is easier than in optical bands. For the
same reasons, the NIR bands are not well-suited for identification of
classical Cepheids.  Moreover, empirical and theoretical
investigations indicate that NIR and optical--NIR PW relations are
independent of metal abundance (\cite[Bono et al. 2010]{bono10};
\cite[Majaess et al. 2011]{majess11}; \cite[Inno et al. 2013]{inno13}).  
To fully exploit these advantages, we performed
a detailed and accurate analysis of NIR and optical--NIR PW relations
(see Table~1).


\begin{table}
  \begin{center}
  \caption{NIR and optical--NIR PW relations for LMC and SMC FU Cepheids.}
  \label{tab1}
 {\scriptsize
  \begin{tabular}{lccccc}\hline 
{$W(\lambda_2,\lambda_1)$$^1$} & {$N_c$} & $a$ & $b$ & {$\sigma$$^2$} &{$\mu$$^3$ (mag)}\\ 
  \hline
\multicolumn{6}{c}{LMC}\\ 
$W(J,K_{\rm s})$ & 1708&  15.876  $\pm$ 0.005  & $-$3.365   $\pm$ 0.008& 0.08   & 18.48 $\pm$ 0.02 \\
$W(J,H)$         & 1701&  15.630  $\pm$ 0.006  & $-$3.373   $\pm$ 0.008& 0.08   & 18.47 $\pm$ 0.02 \\
$W(H,K_{\rm s})$ & 1709&  16.058  $\pm$ 0.006  & $-$3.360   $\pm$ 0.010& 0.10   & 18.50 $\pm$ 0.02 \\
$W(V,K_{\rm s})$ & 1737&  15.901  $\pm$ 0.005  & $-$3.326   $\pm$ 0.008& 0.07   & 18.49 $\pm$ 0.02 \\
$W(V,H)$         & 1730&  15.816  $\pm$ 0.005  & $-$3.315   $\pm$ 0.008& 0.07   & 18.47 $\pm$ 0.02 \\
$W(V,J)$         & 1732&  15.978  $\pm$ 0.006  & $-$3.272   $\pm$ 0.009& 0.08   & 18.47 $\pm$ 0.02 \\
$W(I,K_{\rm s})$ & 1737&  15.902  $\pm$ 0.005  & $-$3.325   $\pm$ 0.008& 0.07   & 18.48 $\pm$ 0.02 \\
$W(I,H)$         & 1734&  15.801  $\pm$ 0.005  & $-$3.317   $\pm$ 0.008& 0.08   & 18.46 $\pm$ 0.02 \\
$W(I,J)$         & 1735&  16.002  $\pm$ 0.007  & $-$3.243   $\pm$ 0.011& 0.10   & 18.44 $\pm$ 0.02 \\
$W(V,I)$         & 1700&  15.899  $\pm$ 0.005  & $-$3.327   $\pm$ 0.008& 0.07   & 18.54 $\pm$ 0.02 \\

{\bf Mean}$^4$ &&&&&  {\bf  18.48 $\pm$ 0.01} \\

\multicolumn{6}{c}{SMC}\\ 
$W(J,K_{\rm s})$ & 2448& 16.457  $\pm$ 0.006 & $-$3.480  $\pm$ 0.011   &0.16  & 18.95 $\pm$ 0.02 \\
$W(J,H)$         & 2448& 16.217  $\pm$ 0.006 & $-$3.542  $\pm$ 0.011   &0.17  & 18.88 $\pm$ 0.02 \\
$W(H,K_{\rm s})$ & 2448& 16.457  $\pm$ 0.006 & $-$3.480  $\pm$ 0.011   &0.19  & 18.99 $\pm$ 0.02 \\
$W(V,K_{\rm s})$ & 2295& 16.507  $\pm$ 0.005 & $-$3.461  $\pm$ 0.011   &0.15  & 18.95 $\pm$ 0.02 \\
$W(V,H)$         & 2285& 16.426  $\pm$ 0.005 & $-$3.475  $\pm$ 0.010   &0.15  & 18.91 $\pm$ 0.02 \\
$W(V,J)$         & 2286& 16.614  $\pm$ 0.005 & $-$3.427  $\pm$ 0.011   &0.16  & 18.94 $\pm$ 0.02 \\
$W(I,K_{\rm s})$ & 2294& 16.511  $\pm$ 0.005 & $-$3.464  $\pm$ 0.011   &0.16  & 18.94 $\pm$ 0.02 \\
$W(I,H)$         & 2202& 16.417  $\pm$ 0.005 & $-$3.480  $\pm$ 0.011   &0.15  & 18.90 $\pm$ 0.02 \\
$W(I,J)$         & 2279& 16.662  $\pm$ 0.006 & $-$3.424  $\pm$ 0.013   &0.18  & 18.91 $\pm$ 0.02 \\
$W(V,I)$         & 2260& 16.482  $\pm$ 0.005 & $-$3.449  $\pm$ 0.010   &0.13  & 18.99 $\pm$ 0.02 \\
{\bf Mean}$^4$ &&&&& {\bf 18.94 $\pm$ 0.01} \\	    
\hline
  \end{tabular}
  }
 \end{center}
\vspace{1mm}
 \scriptsize{Notes:\\
$^1$The color coefficients of the adopted PW relations are
$\frac{A_K}{E(J-K_{\rm s})} =0.69$, 
$\frac{A_H}{E(J-H)}         =1.63$, 
$\frac{A_K}{E(H-K_{\rm s})} =1.92$, 
$\frac{A_K}{E(V-K_{\rm s})} =0.13$, 
$\frac{A_H}{E(V-H)}         =0.22$, 
$\frac{A_J}{E(V-J)}         =0.41$, 
$\frac{A_K}{E(I-K_{\rm s})} =0.24$, 
$\frac{A_H}{E(I-H)}         =0.42$, 
$\frac{A_J}{E(I-J)}         =0.92$, and 
$\frac{A_I}{E(I-V)}         =1.55$.\\
$^2$Standard deviation of the linear fit (mag).\\
$^3$Distance modulus based on the zero-point calibration of S11a.\\
$^4$Weighted distance modulus estimated using the distance moduli of individual PW relations.
}
\end{table}

The results bearing on the linearity of the PW relations deserve a
more detailed discussion. The evidence that optical and NIR PL
relations might not be linear over the entire period range does not
imply that NIR and optical--NIR PW relations have to show the same
trend. The explanation is threefold.
\begin{enumerate}
\item \cite[Bono \& Marconi (1999)]{bono99} suggested that PW
  relations mimic PLC relations. PLC relations, at fixed chemical
  composition, are intrinsically linear, because we correlate the
  pulsation period with both magnitude and color.

\item We can divide the sample into short- ($\log P \lesssim 0.4$
  [days]) and long-period Cepheids, and provide new PW relations for
  the two different subsamples. The use of short- and long-period PW
  relations yields relative MC distances that are less accurate than
  relative distances based on PW relations covering the entire period
  range (Inno et al. 2012).

\item The zero points and slopes of the PW and PLC relations are fixed
  by the most commonly used Cepheids. A glance at the period
  distributions in Fig.~1 shows that they mainly depend on Cepheids
  that are located across the main peaks.  This is the reason why
  short-period PW relations agree quite well with the PW relations
  based on the entire sample. However, the errors in the coefficients
  of the former PW relations are larger than the errors affecting the
  latter (\cite[Inno et al. 2013]{inno13}).
\end{enumerate}

These results relating to the distance to the MCs rely on two
independent calibrations of the zero points. The empirical calibration
relies on trigonometric parallaxes of nine Galactic Cepheids, measured
using the Fine Guidance Sensor on board the {\sl Hubble Space
  Telescope} ({\sl HST}; \cite[Benedict et al. 2007]{benedict07}).
The theoretical calibration relies on different sets of nonlinear,
convective models computed by \cite[Bono et al. (2010)]{bono10} and
\cite[Marconi et al. (2010)]{marconi10}.  To further constrain the
intrinsic accuracy of the zero point of the new PW relations, we
performed a new empirical calibration of 57 Galactic Cepheids. The
main advantage of this sample is that their absolute magnitudes were
estimated using a new calibration of the IRSB method (\cite[Gieren et
  al. 2005]{gieren2005}; \cite[Fouqu\'e et al. 2007]{fouque2007};
\cite[Groenewegen 2008]{groe08}), i.e. a variant of the
Baade--Wesselink method.  Note that the $p$ factor adopted in this
relation to transform the radial velocity into a pulsation velocity
was fixed using the {\sl HST} Cepheids. This means that the current
calibration (see Table~1) is not independent of the calibration based
on {\sl HST} Cepheids. However, we are using a sample of Cepheids that
is a factor of six larger than that composed of {\sl HST} Cepheids and
they cover a broader range in metal abundance.

\begin{figure}[htb!]
\begin{center}
\includegraphics[height=0.60\textheight,width=0.80\textwidth]{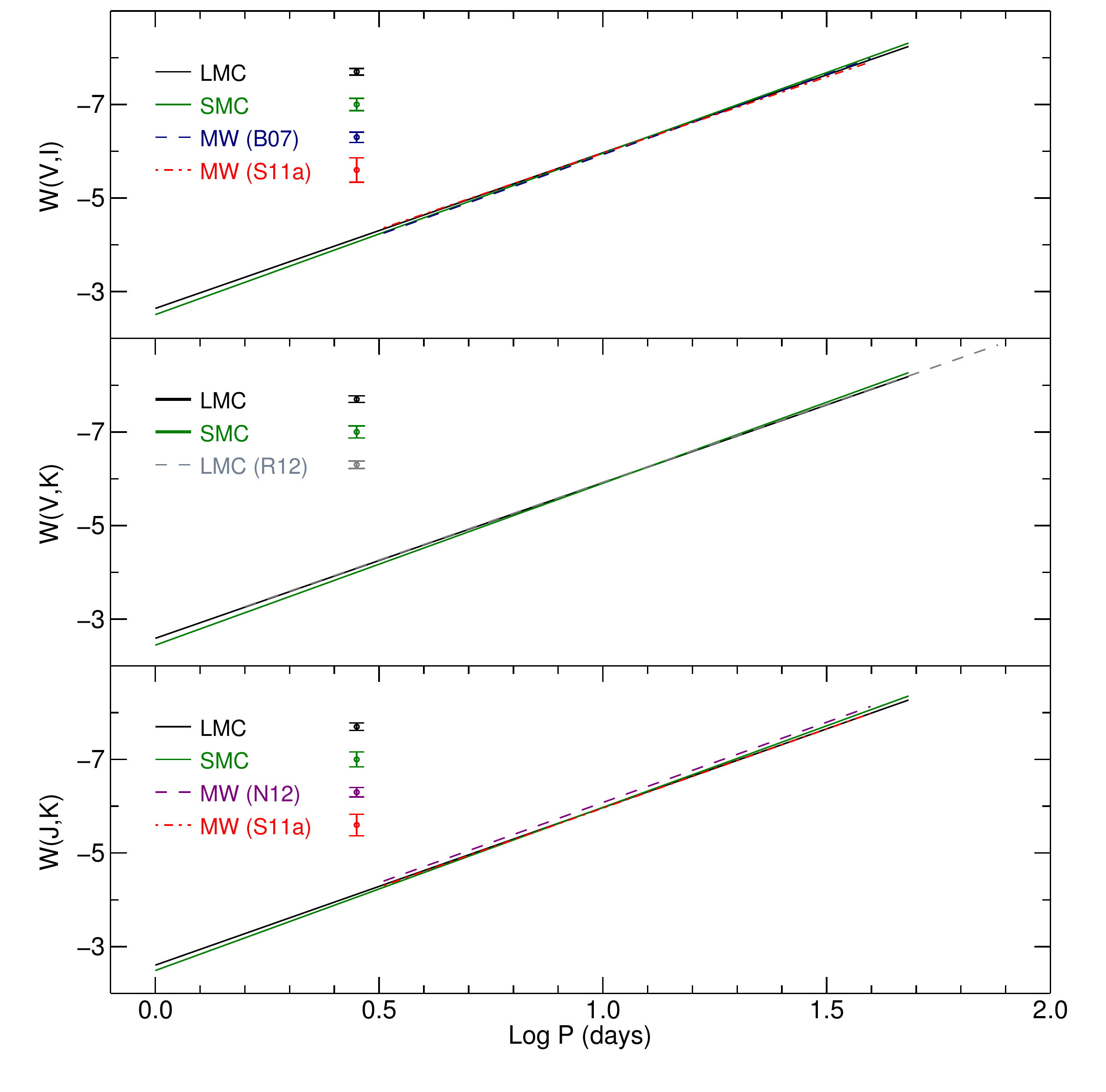}
\vspace*{-0.5 cm}
 \caption{(top) Comparison of current PW $(V,I)$ relations for the LMC
   (black solid line) and SMC (green solid line), and similar PW
   relations for Galactic (MW) Cepheids provided by \cite[Benedict et
     al. (2007]{benedict07}; B07: dashed blue line) and \cite[Storm et
     al. (2011a]{storm}; S11a: dash-dotted red line). The different
   optical--NIR PW relations were plotted for the period range they
   cover. The vertical error bars display the dispersion of the
   different PW relations.  (middle) Same as the top panel, but for
   the PW $(V,K_{\rm s})$ relations.  The PW relation for the LMC was
   provided by \cite[Ripepi et al. (2012]{ripepi12}; R12: grey dashed
   line).  (bottom) Same as the top panel, but for the PW $(J,K_{\rm
     s})$ relations.  The PW $(J,K_{\rm s})$ relations for Galactic
   Cepheids by \cite[Ngeow (2012]{ngeow12}; N12: purple) and S11a
   (red) are also plotted.} \label{fig4}
\end{center}
\end{figure}

Fig.~3 shows a comparison between new NIR and optical--NIR relations
for MC Cepheids and similar relations available in the literature. The
MC and Galactic PW relations plotted in this figure agree quite well
with each other. The difference in the zero points and slopes is
smaller than the dispersion among the individual PW relations (see the
error bars).  Moreover, the new PW relations provide distances to the
MCs that agree quite well with similar estimates available in the
literature (\cite[Inno et al. 2013]{inno13}).


\section{Conclusions and final remarks}

New NIR measurements of MC Cepheids provide the opportunity to
investigate the properties of both optical and NIR PW
relations. Current results indicate that the PW relations are solid
distance indicators, since the slopes, within $1\sigma$ errors, are
independent of metal abundance. No firm conclusion can be reached
regarding the zero points, since we still lack accurate parallaxes for
metal-poor Cepheids. Trigonometric parallaxes are only available for a
few nearby Galactic Cepheids (\cite[Evans et al. 2005]{evans05};
\cite[Benedict et al. 2007]{benedict07}).

The use of MC Cepheids in double-lined spectroscopic binaries appears
a very promising and new, firm opportunity to overcome this
long-standing problem (Pietrzy\'nski et al. and Graczyk et al., this
volume). These systems play a crucial role in constraining thorny
systematic uncertainties affecting the Cepheid distance scale. They
provide not only an independent absolute distance, but also a very
precise measurement of their dynamical mass and radius
(\cite[Pietrzy\'nski et al. 2010, 2011]{pietr10,pietr11}).  This
information allows us to constrain the input physics adopted in
evolutionary and pulsation models.

New optical and NIR surveys of nearby stellar systems provide a
detailed census of regular variables, not only in dwarf irregulars
(\cite[Matsunaga et al. 2011]{matsunaga11}; \cite[Soszy{\'n}ski et
  al. 2012]{sos2012}), dwarf spheroidals (\cite[Pritzl et
  al. 2007]{pritzle}; \cite[Fiorentino et al. 2012]{fiorentino2012}),
dwarf spirals (\cite[Scowcroft et al. 2009]{scowcroft09}), and
ultrafaint dwarfs (\cite[Dall'Ora et al. 2012]{dallora12}), but also
in M31 (\cite[Fliri \& Valls-Gabaud 2012]{fliri12}).  The
observational scenario as regards Galactic Cepheids has also been
improved significantly, and indeed new NIR time-series data provide
the opportunity to identify new Cepheids not only in the nuclear
bulge, but also in the inner disk (Matsunaga, this volume). The NIR
{\sc vvv} survey will also provide a unique opportunity to discover
new variables, and in particular Cepheids, along the Galactic plane
(\cite[Minniti et al. 2010]{minniti10}). Despite these indisputable
new results, we still lack detailed knowledge of the Cepheid
distribution in the outer disk.
   
During the last 10 years, the number of Galactic Cepheids for which we
have iron abundances based on spectroscopic measurements increased
significantly.  Thanks to the investigations by \cite[Luck et
  al. (2011)]{luck11a}, \cite[Luck \& Lambert (2011)]{luck11b},
\cite[Lemasle et al. (2007, 2008)]{lemasle07,lemasle08},
\cite[Romaniello et al. (2008)]{romaniello08}, \cite[Pedicelli et
  al. (2009)]{pedicelli09}, and K. Genovali et al. (in prep.), we have
accurate iron abundances for $\sim$400 Galactic Cepheids.

We already mentioned that metallicity distributions of MC Cepheids
only rely on a few dozen Cepheids. Spectroscopic surveys typically lag
behind photometric surveys. However, the new massively multiplexed
spectrograph (4MOST) for ESO's 4m-class telescopes ({\sl VISTA, NTT})
will certainly play a fundamental role in this context (\cite[de Jong
  2011]{dejong11}; \cite[Ramsay et al. 2011]{ramsay11}). This
instrument will have a field of view in excess of 5 deg$^2$, more than
3000 fibers, and a spectral resolution ranging from 5000 to 20,000,
offering a unique opportunity to map not only the MCs, but also nearby
dwarfs and the Galactic plane.  The same applies to MOONS, the optical
and NIR spectrograph planned for the {\sl VLT} (\cite[Cirasuolo et
  al. 2011]{cirascuolo11}), and to M2FS, the fiber-fed optical
spectrograph at the {\sl Magellan II Clay} telescope (\cite[Mateo et
  al. 2012]{mateo12}).

\section{Acknowledgments}
We are indebted to the editor, R. de Grijs, for his constant support.
One of us (G.B.) thanks ESO for support as a science visitor.  This
work was partially supported by PRIN--INAF 2011, `Tracing the formation
and evolution of the Galactic halo with {\sl VST}' (P.I. M. Marconi)
and by PRIN--MIUR (2010LY5N2T) `Chemical and dynamical evolution of the
Milky Way and Local Group galaxies' (P.I.: F. Matteucci).



\end{document}